\documentclass[12pt,journal,draftclsnofoot,a4paper,oneside,onecolumn]{IEEEtran}

\usepackage{ifpdf}
\usepackage{cite}
\ifCLASSINFOpdf
 \usepackage[pdftex]{graphicx}
\else
  \usepackage[dvips]{graphicx}
\fi
\usepackage[cmex10]{amsmath}
\usepackage{amssymb}
\usepackage{amsbsy}
\usepackage{extarrows}
\usepackage{multirow}

\usepackage{pgf}
\usepackage{pgfplots}
	
\begin{document}

\title{Performance Analysis of Hybrid Relay Selection in Cooperative Wireless Systems}


\author{Tianxi~Liu, Lingyang~Song,~\IEEEmembership{Member,~IEEE,} Yonghui Li,~\IEEEmembership{Senior Member,~IEEE,} Qiang Huo and Bingli~Jiao,~\IEEEmembership{Member,~IEEE}
%
    \thanks{Tianxi~Liu, Lingyang~Song, Qiang~Huo and Bingli~Jiao are with the State Key
Laboratory of Advanced Optical Communication Systems and Networks, School of Electronics Engineering and Computer Science, Peking University, Beijing, China, 100871. e-mail: \protect{tianxi.liu@gmail.com}; \protect{lingyang.song@pku.edu.cn}; \protect{qiang.huo@pku.edu.cn}; \protect{jiaobl@pku.edu.cn}.}
    \thanks{Yonghui~Li is with the University of Sydney, Sydney, NSW 2006, Australia e-mail: \protect{lyh@ieee.org}.}
    }

\maketitle
\begin{abstract}
The hybrid relay selection (HRS) scheme, which adaptively chooses amplify-and-forward~(AF) and decode-and-forward~(DF) protocols, is very effective to achieve robust performance in wireless networks. This paper analyzes the frame error rate (FER) of the HRS scheme in general cooperative wireless networks without and with utilizing error  control coding  at the source node. We first develop an improved signal-to-noise ratio~(SNR) threshold-based FER approximation model. Then, we derive an analytical average FER expression as well as an asymptotic  expression at high SNR for the HRS scheme and generalize to other relaying schemes.  Simulation results are in excellent agreement with the theoretical analysis, which validates the derived FER expressions.
\end{abstract}


\section{Introduction}
\IEEEPARstart{C}ooperative relaying has been shown to be an effective technique to improve the system performance in wireless networks by allowing users to cooperate with each other in their transmissions. Two of the most typical cooperative schemes are amplify-and-forward (AF) and decode-and-forward (DF).
The performance of AF protocol is mainly limited by the noise amplified at the relay during the forwarding process, especially at low SNR.
The performance of DF will be degraded when the relay fails to decode the received signals correctly and the process of decoding and re-encoding will cause serious error propagation. To overcome these limitations, various cooperative schemes have been reported, such as signal-to-noise-ratio (SNR) threshold based selective DF \cite{Laneman2004,Onat2008,Onat2008a,Onat2007}, cooperative-maximum-ratio-combining based DF \cite{Wang2007,Wang2005}, decode-amplify-forward (DAF) \cite{Bao2007,Bao2005DAF}, link adaptive relaying DF \cite{Vien2009,Wang2008,wang2006smart} and log-likelihood-ratio (LLR) threshold based selective DF \cite{kwon2010LLRDF,van2006llr}, etc.

Recently, the hybrid relaying protocol (HRP) has received a lot of attention. It adaptively combines the merits of both DF and AF by forwarding \emph{clean} packets in DF if decodes correctly and forwarding \emph{soft} represented packets in AF if decodes incorrectly) \cite{Can2006DAF-OFDM,Eslamifar2009,Souryal2006,Li2007,Yu2005, Bao2007}.
Its performance can be further improved by incorporating relay selection in HRP, which is referred to as hybrid relay selection (HRS) \cite{Can2006DAF-OFDM,Eslamifar2009,Hasan2009,Li2009,Li2007,Li2008,Song2009,Souryal2006,Yang2009DAF,Yao2009,Yu2005}.
In the HRS scheme, for each transmission, all relays are divided into two groups, a DF group and an AF group. A relay is included either into the DF group if it decodes correctly or into the AF group if it decodes incorrectly. The destination node selects a single optimal relay node with maximum destination SNR from either the DF group or the AF group to forward the packet.

Although the principle of the HRS scheme is simple, the calculation of the analytical FER is non-trivial even for single-relay cooperative networks.  Therefore, most earlier work have been focusing on the approximate average FER analysis. Nonetheless, still very limited results have been reported in the literature so far. In \cite{Souryal2006},  Souryal analyzed the FER of single-relay networks using hybrid forward scheme in block Rayleigh fading channel.
In \cite{Li2007,Li2008}, Li proposed the hybrid relay selection scheme by combining hybrid forwarding and relay selection.
In \cite{Song2009}, Song extended the HRS scheme with differential modulation.
Only very loose upper bounds were provided for the FER of the HRS scheme. 
In \cite{Huo2010P1}, the authors mainly analyze the FER of the all-participate scheme without relay selection, where all relays participate in forwarding signals from the source. 


In this paper, we derive the analytical FER of the HRS scheme in general cooperative wireless networks with and without applying convolutional coding at the source node.
The contribution of this paper is threefold:
1) we develop an improved SNR threshold-based FER approximation model, which is simple and accurate for general diversity systems, and obtain the SNR threshold in an analytical expression;
2) we derive an analytical approximate average FER expression, and its simplified asymptotic FER expression at high SNRs for the HRS scheme;
3) we generalized the derived theoretical FER expressions to other relaying schemes, e.g. the  AF-RS  scheme and the PDF-RS scheme.

The rest of the paper is organized as follows: We describe the system model and the HRS scheme in Section \ref{sec.system.model}. In Section \ref{sec.FER.approx.model}, we develop an improved SNR-threshold based FER approximation model based on cumulative distribution functions (CDF). In Section \ref{sec:FER.HRS}, we
derive the analytical approximate and asymptotic FER expression at high SNRs of the HRS scheme. In Section \ref{sec.simulation}, we present some numerical simulation results. And in Section \ref{sec.conclusion}, we draw the main conclusions.

\emph{\textbf{Notation}}: Boldface lower-case letters denote vectors. $\mathcal{Z}_n$ represents the $n$-dimensional binary space $\left\{0,1\right\}^{n}$. For a random variable $X$, $Pr(\cdot)$ denotes its probability, $f_X(\cdot)$ denotes its probability density function (PDF), $F_X(\cdot)$ is its CDF, and $\mathbb{E}[X]$ represents its expectation. $X\sim \mathcal{CN}(0, \Omega)$ is a circular symmetric complex Gaussian variable with a zero mean and variance $\Omega$. $Q(x)$ denotes the $Q$-function, i.e. $Q(x)=\frac{1}{\sqrt{2\pi}}\int_x^{\infty}e^{-t^2/2}\,\mathrm{d}t$.


\begin{figure}[ht]
\centering
\includegraphics[width=0.8\textwidth]{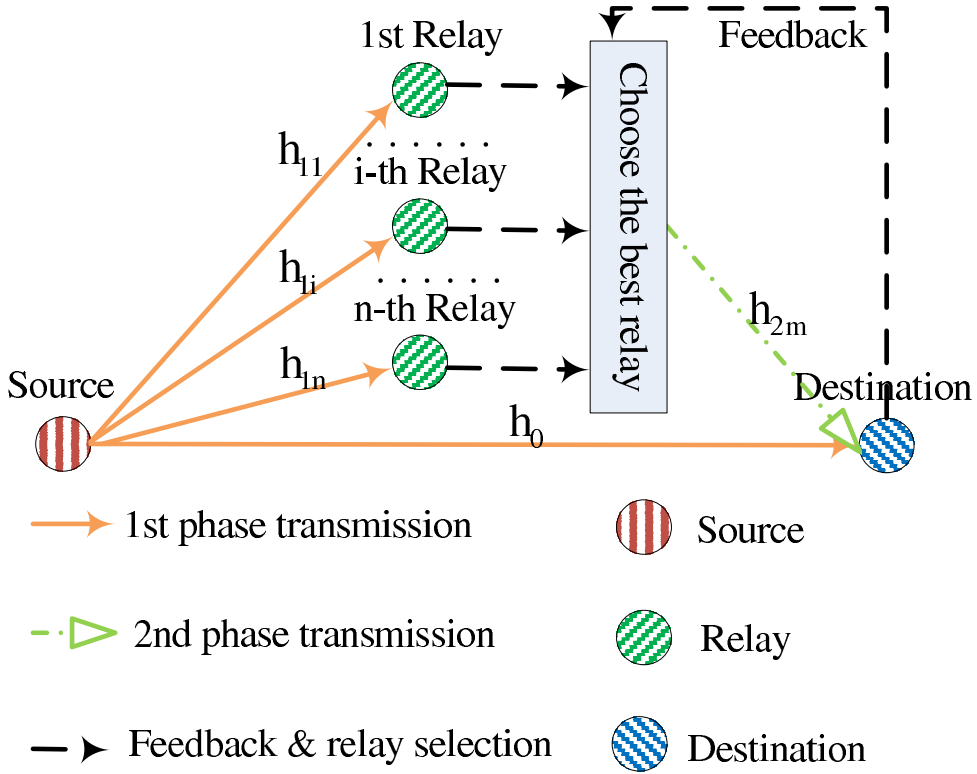}
\caption{Diagram of the cooperative wireless system with the HRS scheme.}  \label{fig:SystemDiagram}
\end{figure}

\section{System Model and The HRS Scheme}\label{sec.system.model}
In this paper, we consider a general $2$-hop relay network, as shown in Fig. \ref{fig:SystemDiagram}, consisting of one source node, $n$ relay nodes and one destination node. The link between any two nodes is modeled as a block Rayleigh fading channel with additive white Gaussian noise~(AWGN), where the fading coefficients of the channels are fixed within one frame and vary independently from one frame to another.

$h_0$, $h_{1,i}$, and $h_{2,i}$ are the fading coefficients of the channels from the source node to the destination node, from the source node to the $i$-th relay node, and from the $i$-th relay node to the destination node, respectively. We assume $h_0\sim \mathcal{CN}(0,\Omega_0)$, $h_{1i} \sim\mathcal{CN}(0,\Omega_{1i})$, and $h_{2,i} \sim\mathcal{CN}(0,\Omega_{2i})$. Similarly, $n_0$, $n_{1i}$ and $n_{2i}$ are the corresponding additive Gaussian noises. We assume $n_0 \sim \mathcal{CN}(0,N_0)$, $n_{1i} \sim \mathcal{CN}(0,N_0)$, and $n_{2i} \sim \mathcal{CN}(0,N_0)$. Without loss of generality, we consider that all the nodes transmit with the same power
$\mathcal{E}$. In addition, all channel state information (CSI) needed for decoding is available at the relay nodes and the destination node. 

We assume that each terminal in the network is equipped with a single antenna working in the half-duplex mode. We consider an orthogonal transmission scheme in which only one terminal is allowed to transmit at each time slot. Therefore, one frame transmission in the HRS scheme consists of two phases:

\subsection{The First Phase}
The source node broadcasts the transmit signals with length $L$, denoted by $\textbf{s}$, to both the destination node and all of the relay nodes. The received signals at the destination node and the $i$-th relay
node, denoted by $\textbf{y}_0$ and $\textbf{y}_{1i}$, are $ \textbf{y}_0 = \sqrt{\mathcal{E}}h_0 \textbf{s} + \textbf{n}_0, $ and $ \textbf{y}_{1i} = \sqrt{\mathcal{E}} h_{1i} \textbf{s} +\textbf{n}_{1i}, $ respectively. Then the corresponding instantaneous SNRs
can be calculated as
\begin{equation*} \label{eq:SNR_S2D}
\gamma_0 = \frac{\mathbb{E}(\lvert \sqrt{\mathcal{E}}
h_0 \textbf{s}  \rvert ^2)}{\mathbb{E}(\lvert \textbf{n}_0 \rvert ^2)}=
\frac{\mathcal{E}}{N_0} |h_0|^2 = \bar{\gamma} |h_0|^2 ,
\end{equation*}
and
\begin{equation*} \label{eq:SNR_S2Ri}
\gamma_{1i} = \frac{ \mathcal{E} }{N_0}|h_{1i}|^2 = \bar{\gamma}
|h_{1i}|^2,
\end{equation*}
respectively, where $\mathbb{E}(\lvert \mathbf{s}\rvert ^2)=L$,  $\mathbb{E}(\lvert \mathbf{n}\rvert ^2)=LN_0$, and $\bar{\gamma} $ is defined as
\begin{equation*}
\bar{\gamma} \triangleq\frac{\mathcal{E}}{N_0}.
\end{equation*}

\subsection{The Second Phase}
In the second phase, every relay node first decodes it's receive signals. Then it is assigned to either the AF group $\mathcal{G}_{AF}$ or the DF group $\mathcal{G}_{DF}$ according to its CRC checking result. If the CRC checking result is correct, the relay node is included into $\mathcal{G}_{DF}$, otherwise, it is included into $\mathcal{G}_{AF}$. As a result, the instantaneous destination SNR of the link through the $i$-th relay node, denoted by $\gamma_i$, can be expressed as
\begin{equation} \label{eq:SNR_i_cases}
\begin{split}
\gamma_i &= \begin{cases}
\frac{\gamma_{1i}\gamma_{2i}}{\gamma_{1i}+\gamma_{2i}+1}, & \text{if $i \in \mathcal{G}_{AF}$ }, \\
\gamma_{2i}, & \text{if $ i \in \mathcal{G}_{DF}$ },
\end{cases} \\
\end{split}
\end{equation}
where $\gamma_{2i}$ represents the instantaneous SNR of the link between the $i$-th relay node and the destination node, which is given by
\begin{equation*}
\gamma_{2i} = \frac{ \mathcal{E} }{N_0} |h_{2i}|^2 = \bar{\gamma}
|h_{2i}|^2.
\end{equation*}

Then, the destination node selects the relay node with maximum SNR $\gamma_m$ to transmit at the second phase
\begin{equation} \label{eq:Selection}
\gamma_m = \max_{1 \leq i \leq n} \{ \gamma_i \}.
\end{equation}

The selected relay node $m$ transmits through AF if $m \in \mathcal{G_{AF}}$ and using DF if $m \in \mathcal{G_{DF}}$. Finally, the destination combines the signals received at both phases by maximal ratio combining (MRC) to decode. After MRC, the effective SNR of the received signal, denoted by $\gamma_{HRS}$, is calculated as
\begin{equation} \label{eq:gamma_vz}
\gamma_{HRS} = \gamma_{0} + \gamma_m.
\end{equation}

The selection process in the HRS scheme is exactly the same as in the conventional AF or DF selection schemes and does not add any extra complexity in system implementation. The only requirement for the HRS scheme is that each relay needs send one bit indicator to inform the destination if it uses AF or DF protocol. Destination then calculates the overall received SNR from each relay accordingly and selects the best relay with largest SNR.

For simplicity, we define $\lambda_{0}$, $\lambda_{1i}$ and $\lambda_{2i}$ as:
\begin{equation} \label{eq.lambda.define}
\lambda_{0} = \frac{1}{\bar{\gamma} \Omega_{0}}, \lambda_{1i} = \frac{1}{\bar{\gamma} \Omega_{1i}}, \lambda_{2i} = \frac{1}{\bar{\gamma} \Omega_{2i}}.
\end{equation}

\section{An Improved SNR Threshold-Based FER Approximation Model} \label{sec.FER.approx.model}

In this section, we develop an improved SNR threshold-based FER approximation model for general diversity systems. 

We first describe the SNR threshold-based FER approximation model in Subsection \ref{subsec.FER.model.introduction}, in Subsection \ref{subsec.previous.SNR}, some earlier  approaches are presented, and then we propose an improved criterion to calculate the SNR threshold in Subsection \ref{subsec.improved.SNR}.

\subsection{Introduction of the SNR threshold-based FER approximation model} \label{subsec.FER.model.introduction}

The average FER over a block fading channel, denoted by $\bar{P}_f$, can be computed by integrating the instantaneous FER over AWGN channel, represented by $P_f^G(\gamma)$, over the fading distribution \cite{Proakis1995}
\begin{equation} \label{eq:P_f_B}
\bar{P}_{f}(\bar{\gamma})= \int_{0}^{\infty} P_f^G(\gamma) f_{\gamma}(\gamma,\bar{\gamma})
\text{d} \gamma,
\end{equation}
where $\gamma$ and $\bar{\gamma}$ denote the instantaneous  and average SNR, and $f_{\gamma}(\cdot)$ denotes
the PDF of $\gamma$.

Although (\ref{eq:P_f_B}) is an exact expression for $\bar{P}_f $, its closed-form expression is difficult to evaluate. By assuming the instantaneous FER is $1$ when instantaneous SNR $\gamma$ is below a threshold $\gamma_t$, otherwise it is $0$:
\begin{equation*}
P_f^G(\gamma|\gamma \leq \gamma_t) \approx 1 \; \text{and} \;
P_f^G(\gamma|\gamma > \gamma_t) \approx 0 ,
\end{equation*}
the average FER can be rewritten as
\begin{equation} \begin{split} \label{eq:ThresholdModelFER}
\bar{P}_f &= \int_{0}^{\gamma_t} P_f^G(\gamma) f_{\gamma}(\gamma,\bar{\gamma})
\text{d} \gamma
+ \int_{\gamma_t}^{\infty} P_f^G(\gamma) f_{\gamma}(\gamma,\bar{\gamma}) \text{d} \gamma \\
&\approx \int_{0}^{\gamma_t} f_{\gamma}(\gamma,\bar{\gamma}) \text{d} \gamma =
F_{\gamma}(\gamma_t,\bar{\gamma}),
\end{split} \end{equation}
where $f_{\gamma}(\cdot)$ and $F_{\gamma}(\cdot)$ are the PDF and CDF of $\gamma$, respectively. 

Note that, according to the approximation model, the analytical FER can be calculated as an outage probability, and thus, the accuracy is mainly determined by the SNR threshold selection.

\subsection{Some Existing SNR Threshold Approaches}\label{subsec.previous.SNR}

In \cite{ElGamal2001}, El Gamal and Hammons demonstrated the SNR threshold-based FER approximation model for iteratively decoded systems employing turbo codes. And the optimal SNR threshold has been proved to coincide with the convergence threshold of the iterative turbo decoder. 

Recently, in \cite{Chatzigeorgiou2008,Chatzigeorgiou2009},  Chatzigeorgiou extended this model to non-iterative coded and uncoded systems. To get the optimal SNR threshold, proper error criterion should be used. In \cite{Chatzigeorgiou2008,Chatzigeorgiou2009}, the \emph{minimum absolute error sum criterion} is adopted to minimize the sum of absolute error
\begin{equation*} \label{eq:AbsoluteCriterion}
\gamma_{t} = \text{arg} \; \min\left\{ \int_{0}^{\infty} \left|
\bar{P}_{f}(\bar{\gamma}) - F_{\gamma}(\gamma_t,\bar{\gamma})\right|
\text{d} \bar{\gamma} \right\} \;,
\end{equation*}
where $\bar{\gamma}$ is the average SNR, and the SNR threshold can be calculated as
\begin{equation} \label{eq.SNR.threshod.old}
\gamma_t = \left(\int_{0}^{\infty} \frac{1-P_f^G(\gamma)}{\gamma^2}
\text{d} \gamma \right)^{-1}.
\end{equation}

However, the model on the basis of the \emph{minimum absolute error sum criterion} might not be sufficiently accurate since it does not consider the fact that FER decreases more quickly at high SNR region in high diversity order systems. Hence, it can be improved.

\subsection{Model improvement and the SNR threshold result} \label{subsec.improved.SNR}

In this subsection, we propose an improved SNR threshold-based FER approximation model for the general diversity systems. The improvements are twofold.

Firstly, taking into account the fact that the FER decreases quickly when SNR increases, \emph{minimum absolute relative error sum criterion}, which minimizing the sum of absolute \emph{relative} error, can be adopted
\begin{equation} \label{eq:RelativeCriterion}
\gamma_{t} = \text{arg} \; \min\left\{ \int_{0}^{\infty} \left|
\frac{ \bar{P}_{f}(\bar{\gamma}) -
F_{\gamma}(\gamma_t,\bar{\gamma})}{\bar{P}_{f}(\bar{\gamma})}\right|
\text{d} \bar{\gamma} \right\} \; .
\end{equation}

As Eq. (\ref{eq:RelativeCriterion}) is difficult to solve, we instead use a suboptimal error criterion. Reasonably, suppose $\min\left\{ \int_{0}^{\infty} \left|
\frac{ \bar{P}_{f}(\bar{\gamma}) -
F_{\gamma}(\gamma_t,\bar{\gamma})}{\bar{P}_{f}(\bar{\gamma})}\right|
\text{d} \bar{\gamma} \right\} < \infty$ and notice that the integration in Eq. (\ref{eq:RelativeCriterion}) is from $0$ to $\infty$, and then, the absolute relative error should approach zero when $\bar{\gamma} \rightarrow \infty$:
\begin{equation}
 \begin{split}
\label{eq:ZECriterion} \gamma_{t}
&= \text{arg} \lim_{\bar{\gamma}
\rightarrow \infty} \left\{
\left|
\frac{ \bar{P}_{f}(\bar{\gamma}) -
F_{\gamma}(\gamma_t,\bar{\gamma})}{\bar{P}_{f}(\bar{\gamma})}\right|
=0 \right\} \\
& \approx \text{arg} \lim_{\bar{\gamma}
\rightarrow \infty} \left\{\bar{P}_{f}(\bar{\gamma})
-F_{\gamma}(\gamma_t,\bar{\gamma}) =0 \right\}.
 \end{split}
\end{equation}
Otherwise, for a sufficiently big value $T$ ($0<T<\infty$) and a small enough value $\delta$ ($0< \delta < \infty$), the absolute relative error can be greater than $\delta$, i.e. $\left|
\frac{ \bar{P}_{f}(\bar{\gamma}) -
F_{\gamma}(\gamma_t,\bar{\gamma})}{\bar{P}_{f}(\bar{\gamma})}\right| >\delta, \text{when } \bar{\gamma}>T$.  Hence, the absolute relative error sum can't be minimized as it approaches infinity: $\int_{0}^{\infty} \left|
\frac{ \bar{P}_{f}(\bar{\gamma}) -
F_{\gamma}(\gamma_t,\bar{\gamma})}{\bar{P}_{f}(\bar{\gamma})}\right|
\text{d} \bar{\gamma} > \int_{T}^{\infty} \delta \text{d} \bar{\gamma} = \infty$.

Secondly, the SNR threshold should include the factor of the diversity order of general diversity systems.  For example, in a Rayleigh fading channel, the CDF of SNR is $F_{\gamma}(\gamma, \bar{\gamma}) = 1- e^{-\gamma / \bar{\gamma}}$, then $\lim_{\bar{\gamma} \rightarrow \infty }F_{\gamma}(\gamma, \bar{\gamma})
< {\bar{\gamma}}^{-1} \gamma $ and in a Nakagami-$m$ fading channel, the CDF of SNR is given by: $F_{\gamma}(\gamma, \bar{\gamma}) = \frac{\Gamma(m,m\gamma/\bar{\gamma})}{\Gamma(m)}$, where $\Gamma(m)$ is the Gamma function, and $\Gamma(m,x)$ is the lower part incomplete Gamma function given by: $\Gamma(m,x)=\int_{0}^x t^{m-1}e^{-t}\text{d}t$, then using the fact that $\lim_{t\rightarrow 0}e^{-t}<1$, we can get $\lim_{\bar{\gamma} \rightarrow \infty }F_{\gamma}(\gamma, \bar{\gamma})
< \Gamma(m)^{-1}\int_{0}^{\bar{\gamma}} (\frac{\gamma}{\bar{\gamma}})^{m-1}\text{d}t =\frac{1}{m \Gamma(m) (\bar{\gamma})^m } \gamma^m $.

Suppose for some diversity system with a diversity order of $d$, 
the CDF of SNR can be approximated in a form \cite{Chatzigeorgiou2009,Zheng2003,Rodrigues2008Turbo}:
\begin{equation} \label{eq:CDFLimit}
\lim_{\bar{\gamma} \rightarrow \infty }F_{\gamma}(\gamma, \bar{\gamma})
\approx G(\bar{\gamma}) \gamma ^d,
\end{equation}
where $G(\bar{\gamma})$ is a constant related to $\bar{\gamma}$.

Then, combining Eq. (\ref{eq:P_f_B}), Eq. (\ref{eq:ZECriterion}) and Eq. (\ref{eq:CDFLimit}), we can get the analytical SNR threshold for our improved model as
\begin{equation} \label{eq:SNR_threshold_continue}
\begin{split}
\gamma_{t,d} &\approx \text{arg} \lim_{\bar{\gamma} \rightarrow \infty}
\left\{ \int_{0}^{\infty} f_{\gamma}(\gamma, \bar{\gamma})
P_f^G(\gamma) \text{d} \gamma
- G(\bar{\gamma}) \gamma_t ^d= 0\right\} \\
&=\text{arg} \lim_{\bar{\gamma} \rightarrow \infty} \left\{
\int_{0}^{\infty} G(\bar{\gamma}) d \gamma ^{d-1} P_f^G(\gamma)
\text{d} \gamma - G(\bar{\gamma}) \gamma_t ^d
= 0\right\} \\
&= \text{arg} \lim_{\bar{\gamma} \rightarrow \infty} \left\{ {\gamma_t}^d = d \int_{0}^{\infty} \gamma^{d-1} P_f^G(\gamma) \text{d} \gamma \right\} \\
&\approx \left( d \int_{0}^{\infty} \gamma^{d-1} P_f^G(\gamma) \text{d}
\gamma \right)^{1/d}.
\end{split}\end{equation}

 In  uncoded systems, $ P_f^G(\gamma) $ can be given in closed-form. For example, for linear modulation, $ P_f^G(\gamma) $ can be approximated  as \cite{Chatzigeorgiou2008}
\begin{equation} \label{eq:P_f_G_uncode}
P_f^G(\gamma) \approx 1 - \left(1 - Q\left(\sqrt{c \gamma}\right)\right)^L,
\end{equation}
where $c$ is modulation constant ($c=2$ for binary-phase-shift-keying (BPSK)), $Q(\cdot)$ is the $Q$ function, $L$ is the frame length. Substituting Eq. (\ref{eq:P_f_G_uncode}) into Eq. (\ref{eq:SNR_threshold_continue}), we can get SNR threshold $\gamma_{t,d}$ for uncoded systems:
\begin{equation} \label{eq:SNR_threshold_uncoded}
\gamma_{t,d} \approx  \left( d \int_{0}^{\infty} \gamma^{d-1} \left(1 - \left(1 - Q\left(\sqrt{c \gamma}\right)\right)^L \right)  \text{d}
\gamma \right)^{1/d}.
\end{equation}

In coded systems, we can calculate Eq. (\ref{eq:SNR_threshold_continue}) using numerical methods. We can first get $P_f^G(\gamma) $, i.e. the instantaneous FER for the scheme over AWGN channel, using Monte Carlo methods. When the SNR values $\gamma_i^{\prime}$, $i=1,2,\cdots,N$, are equally spaced with $\Delta \gamma$ and ordered, the following equivalent expression for discrete SNR values can be obtained
\begin{equation}\label{eq:SNR_threshold_quanti}
\gamma_{t,d} \approx \left( d \sum_{i=1}^{N}\gamma_i^{\prime d-1} P_f^G(\gamma_i^{\prime})
\Delta \gamma\right)^{1/d} .
\end{equation}

Note that when $c=2$ (BPSK), $L=1$, and $d=1$, we can obtain $\gamma_{t,1}=\frac{1}{4}$ using Eq. (\ref{eq:SNR_threshold_uncoded}) \cite{GammaBetaErf}, and $\bar{P}_f = 1-e^{-\frac{1}{4\bar{\gamma}}} \approx \frac{1}{4\bar{\gamma}}$, which is equivalent to the well known average BER of BPSK over Rayleigh fading channel at high SNR: $\bar{P}_b=\frac{1}{2}(1-\frac{\bar{\gamma}}{1+\bar{\gamma}}) \approx \frac{1}{4\bar{\gamma}}$ \cite{Proakis1995}. 

If we consider a multiple-input multiple-output (MIMO) channel having $N_T$ inputs and $N_R$ outputs. The transmitter uses space-time block coding \cite{alamouti1998simple}, while the receiver coherently combines the $N=N_T N_R$ independent fading paths. If $\gamma$ now corresponds to the instantaneous SNR at the output of the combiner, its probability distribution is given by \cite{Proakis1995,Chatzigeorgiou2009}:
\begin{equation*}
f_{\bar{\gamma}}(\gamma) = \frac{\gamma^{N-1}e^{-\gamma/(\bar{\gamma}/N_T)}}{(\bar{\gamma}/N_T)^N (N-1)!},
\end{equation*}
where $\bar{\gamma}$ is the average SNR per receive antenna. Using the fact that $e^{-x} < 1$ ($x>0$), the CDF of $\gamma$ can be approximated as:
\begin{equation*}
 \begin{split}
F_{\bar{\gamma}}(\gamma) &\approx \int_0^{\gamma} \frac{t^{N-1}}{(\bar{\gamma}/N_T)^N (N-1)!} \text{d} t\\
&=\frac{1}{(\bar{\gamma}/N_T)^N N!} \gamma^{N}\\
&
=G(\bar{\gamma}) \gamma^{N},
\end{split}
\end{equation*}
where $G(\bar{\gamma}) =\frac{1}{(\bar{\gamma}/N_T)^N N!}$. So, we can obtain the SNR threshold $\gamma_{t,N}$ using above method, and the approximated
FER of the system for MIMO quasistatic fading channels can then be
approximated as \cite{Proakis1995,Chatzigeorgiou2009}
\begin{equation*}
\begin{split}
\bar{P}_f &\approx F_{\bar{\gamma}}(\gamma_{t,N}) \\
&\approx 1- e^{-\gamma_{t,N}N_T/\bar{\gamma}}\sum_{k=0}^{N-1}\frac{(\gamma_{t,N} N_T /\bar{\gamma})^k}{k!} \\
&< \frac{1}{(\bar{\gamma}/N_T)^N N!} \gamma_{t,N}^{N}.
\end{split}
\end{equation*}

So far, we have developed an improved SNR threshold-based FER approximation model
for general diversity systems. In this model, FER is approximated as an
outage probability as Eq. (\ref{eq:ThresholdModelFER}), the SNR threshold is given by Eq. (\ref{eq:SNR_threshold_continue}) and can be calculated using Eq. (\ref{eq:SNR_threshold_uncoded})  for uncoded systems and  using Eq. (\ref{eq:SNR_threshold_quanti}) for coded systems.



\section{FER Analysis of the HRS Scheme}\label{sec:FER.HRS}

 In this section, we perform the FER analysis of the HRS scheme in cooperative systems.


\subsection{Analytical FER Analysis of the HRS Scheme}\label{subsec:FER_HRS}

According to the proposed FER approximation model, a frame error only occurs if SNR is below the SNR threshold $\gamma_{t,d}$. Then, the average FER of
the HRS scheme, denoted by $\bar{P}_f$, can be expressed as
\begin{equation} \label{eq:Pf} \begin{split}
\bar{P}_f &=   Pr({\gamma_{HRS}} < \gamma_{t,d}) ,
\end{split} \end{equation}
where $\gamma_{HRS}$ is the SNR of the HRS scheme  and $d$ is the diversity order of the cooperative system with the HRS scheme.

Hence, to get the FER for the HRS scheme, we merely need to derive the CDF of  $\gamma_{HRS}$. Since the channels are Rayleigh block fading, the instantaneous SNR $\gamma_{1i}$ is exponentially distributed. Then, the CDF of $\gamma_{1i}$ is given by
\begin{equation*} \label{eq:CDF_SNR_{1i}}
F_{\gamma_{1i}}(\gamma_{t,1}) = 1 - e^{-\lambda_{1i} \gamma_{t,1} },
\end{equation*}
where $\lambda_{1i}=1/\gamma_{1i}$ is defined in Eq.  (\ref{eq.lambda.define}).

For simplicity, we introduce a vector variable $\mathbf{z}$, where
\begin{equation*}
\mathbf{z}=\left[z_1, z_2, \cdots, z_n\right], z_i = \begin{cases}
0, & \text{if $\gamma_{1i} \geq \gamma_{t,1}$}, \\
1, & \text{if $\gamma_{1i}< \gamma_{t,1}$}.
\end{cases}
\end{equation*}

In the HRS scheme, the $i$-th relay node is assigned to either $\mathcal{G}_{DF}$ or $\mathcal{G}_{AF}$ according to its SNR $\gamma_{1i}$. If $\gamma_{1i} \geq \gamma_{t,1}$ ($z_i = 0$) then $i \in \mathcal{G}_{DF}$, and if $\gamma_{1i} < \gamma_{t,1}$ ($z_i = 1$) then $i
\in \mathcal{G}_{AF}$. By using the fact that for any value $x$, $x^0=1$, then the probability of $z_i$, denoted by $Pr(z_i)$, can be written as
\begin{equation*} \begin{split}
Pr(z_i)
&= [Pr(z_i=0)]^{1-z_i}[ Pr(z_i=1) ]^{z_i}\\
&= [Pr(\gamma_{1i} \geq \gamma_{t,1})]^{1-z_i}[ Pr(\gamma_{1i} < \gamma_{t,1}) ]^{z_i}\\
&=[ F_{\gamma_{1i}}(\gamma_{t,1}) ]^{z_i} [ 1- F_{\gamma_{1i}}(\gamma_{t,1})
]^{1 - z_i}.
\end{split} \end{equation*}


Then, the CDF of $\gamma_{HRS}$ is derived in Appendix
\ref{Proof:CDF_SNR_Z} as
\begin{equation} \label{eq:CDF_SNR_Z}
F_{\gamma_{HRS}}(\gamma_{t,d}) \approx \sum_{\mathbf{b} \in \mathcal{Z}_n, \mathbf{b}\neq0}
\mathcal{C}_1
\frac{\mathcal{C}_2(1-e^{- \lambda_{0} \gamma_{t,d}})-\lambda_0(1-e ^{-
\mathcal{C}_2 \gamma_{t,d}}) }{\lambda_0-\mathcal{C}_2},
\end{equation}
where $\mathbf{b}=\{b_i, i=1,\cdots,n\} \in \mathcal{Z}_n, b_i \in \{0,1\}$, $\mathcal{C}_1 = (-1)^{\sum_{i=1}^n
b_i} $ and $
\mathcal{C}_2 = \sum_{i=1}^n b_i (\frac{\gamma_{t,1}}{\gamma_{t,d}}\lambda_{1i} + \lambda_{2i})$.

Substituting   Eq. (\ref{eq:CDF_SNR_Z}) into Eq. (\ref{eq:Pf}), we can obtain the FER for the HRS scheme
\begin{equation} \begin{split} \label{eq:FER_HRS}
\bar{P}_{f} \approx \sum_{\mathbf{b} \in \mathcal{Z}_n, \mathbf{b}\neq0} \mathcal{C}_1
\frac{\mathcal{C}_2(1-e^{- \lambda_{0} \gamma_{t,d}})-\lambda_0(1-e ^{-
\mathcal{C}_2 \gamma_{t,d}}) }{\lambda_0-\mathcal{C}_2}.
\end{split} \end{equation}


\subsection{FER Simplification at High SNR}
\label{subsec:FER_Simplify}
In this subsection, we derive the simplified FER for the HRS scheme at high SNR. At high SNR, $\bar{\gamma_0}$, $\bar{\gamma_{1i}}$ and $\bar{\gamma_{2i}}$ are large, then according to Eq. (\ref{eq.lambda.define}):  $\lambda_0=1/\bar{\gamma_0}$, $\lambda_{1i}=1/\bar{\gamma_{1i}}$ and $\lambda_{2i}=1/\bar{\gamma_{2i}}$ are small. Using the approximation that $ 1- \exp(- x) \approx x, \text{when $x \rightarrow 0$}, $ we can get
the simplified CDF of $\gamma_{HRS}$ at high SNR is given by
\begin{equation} \label{eq:CDF_SNR_Z_approx}
F_{\gamma_{HRS}} (\gamma_{t,d}) \approx  \frac{\lambda_0 \gamma_{t,d}^{n+1}}{n+1} \prod_{i=1}^n
(\frac{\gamma_{t,1}}{\gamma_{t,d}}\lambda_{1i} + \lambda_{2i}).
\end{equation}
The proof of Eq. (\ref{eq:CDF_SNR_Z_approx}) is given at appendix \ref{Proof:CDF_SNR_Z_approx}.

Substituting    Eq. (\ref{eq:CDF_SNR_Z_approx}) into Eq. (\ref{eq:Pf}), we can obtain the simplified FER of the HRS scheme at high SNR
\begin{equation} \begin{split}  \label{eq:FER_HRS.simplify}
\bar{P}_{f} &\approx
\frac{(\gamma_{t,d})^{n+1}}{n+1} \lambda_0
 \prod_{i=1}^n  \lambda_{2i} \left(1+ \frac{\gamma_{t,1}}{\gamma_{t,d}} \frac{\lambda_{1i}}{ \lambda_{2i}}\right),
\end{split} \end{equation}
where $\lambda_0  \prod_{i=1}^n \left(\lambda_{1i} + \lambda_{2i}\right) \propto (\bar{\gamma})^{-(n+1)}$, which indicates that the diversity order of $n+1$ can be achieved, $d=n+1$,  $\gamma_{t,d}=\gamma_{t,n+1}$ and see also Eq. (\ref{eq.binomial}) for the last step.

The simplified FER is more intuitive and simple compared to the  FER expression in Eq. (\ref{eq:FER_HRS}). Note that both Eq. (\ref{eq:FER_HRS}) and Eq. (\ref{eq:FER_HRS.simplify}) are analytical approximate average FER expressions for the HRS scheme in cooperative networks. The calculation methods of parameters $\gamma_{t,1}$ and $\gamma_{t,d}$ are presented in Section \ref{sec.FER.approx.model}.

\subsection{Extending to other Relaying Schemes} \label{subsec.AF.PDF}

The FER results for the HRS scheme in  Eq. (\ref{eq:FER_HRS}) and Eq. (\ref{eq:FER_HRS.simplify}) can be easily  extended to the cooperative systems using the AF with relay selection (AF-RS) scheme \cite{Zhao2006} and the perfect DF with relay selection (PDF-RS) scheme \cite{Beres2006}. In the AF-RS scheme, a single relay with the maximum source-relay-destination SNR will be selected to forward signals from the source to the destination using AF  \cite{Zhao2006}. In the PDF-RS scheme, it assumes that all relay nodes can correctly decode the received signals and a single relay with the maximum relay-destination SNR will be selected to forward signals from the source to the destination. For simplicity, we only present the simplified results at high SNR  \cite{Beres2006}.

For PDF-RS, the transmissions from the source to any relay are always correct, it means that $\gamma_{1i}=\infty$ and $z_i=0$.  Thus, the  average FER of PDF-RS, denoted by $\bar{P}_f^{PDF-RS}$, is given by
\begin{equation*}\label{eq.FER.PDF-RS}
\bar{P}_f^{PDF-RS}\approx \frac{(\gamma_{t,d})^{n+1}}{n+1} \lambda_0
 \prod_{i=1}^n  \lambda_{2i}.
\end{equation*}

For AF-RS, the SNR of the $i$-th Rayleigh relay link (from the source to the $i$-th Rayleigh relay and to the destination) can be approximated as "combined" Rayleigh fading with parameter of $\lambda_{i}^{AF}\approx\lambda_{1i}+\lambda_{2i}$ at high SNR \footnote{This is a lower bound for AF \cite{Anghel2004,LiuTianxi2010}.}. Using this approximation, AF-RS can be viewed as a special case of PDF-RS with $\lambda_{2i}^{'}=\lambda_{i}^{AF}$. The average FER of AF-RS, denoted by $\bar{P}_f^{AF-RS}$, is given by
\begin{equation*} \label{eq.FER.AF-RS}
\bar{P}_f^{AF-RS}\approx \frac{(\gamma_{t,d})^{n+1}}{n+1} \lambda_0
 \prod_{i=1}^n  \lambda_{2i} \left(1+ \frac{\lambda_{1i}}{ \lambda_{2i}}\right).
\end{equation*}


Defining the gain of the HRS scheme over the AF-RS scheme as the ratio of FER, we can get
\begin{equation*}\label{eq.gain.HRS.AF-RS}
G_{\frac{HRS}{AF-RS}} = \frac{\bar{P}_f^{AF-RS}}{\bar{P}_{f} } \approx \prod_{i=1}^n  \frac{ 1+ \frac{\lambda_{1i}}{ \lambda_{2i}}}{1+ \frac{\gamma_{t,1}}{\gamma_{t,d}} \frac{\lambda_{1i}}{ \lambda_{2i}} } \text{(dB)}.
\end{equation*}
It indicates that the gain improves when $\frac{\bar{\gamma}_{2i}}{ \bar{\gamma}_{1i}}$ increases, and when the number of relay nodes increases, the gain increases quickly.


\section{Simulation results} \label{sec.simulation}

\begin{figure}[!t]
\centering
\includegraphics[width=0.8\textwidth]{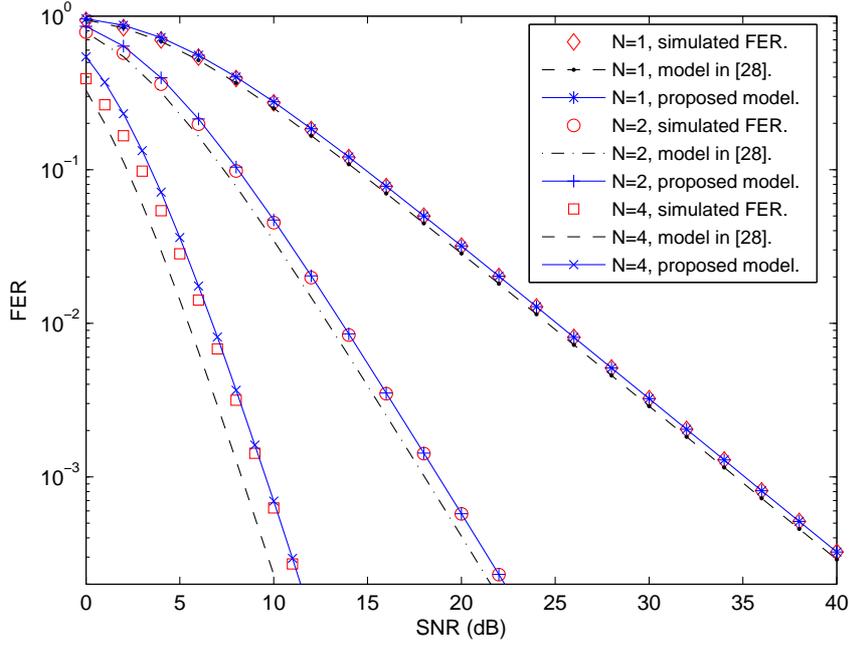}
\caption{FER comparison of proposed model and the model of  \cite{Chatzigeorgiou2008} for case 0: general MIMO channels with $N_T=1$, $N=N_R=1,2,4$, uncoded.} \label{fig.case.0}
\end{figure}



\begin{figure}[!t]
\centering
\includegraphics[width=0.8\textwidth]{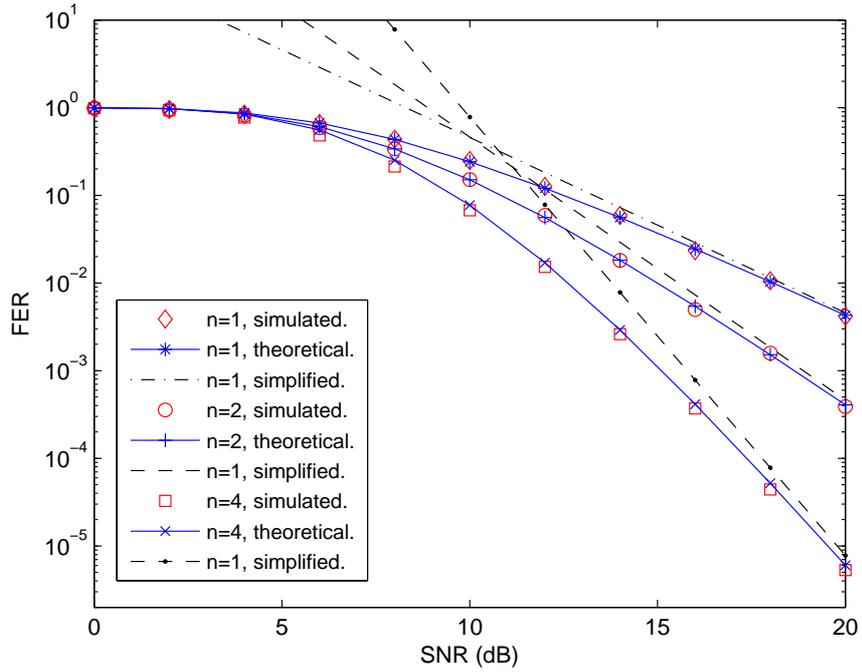}
\caption{Average FER of the HRS scheme for case 1: $\Omega_0=\Omega_{1i}=\Omega_{2i}=1$, uncoded.} \label{fig.case.1}
\end{figure}

%


\begin{figure}[!t]
\centering
\includegraphics[width=0.8\textwidth]{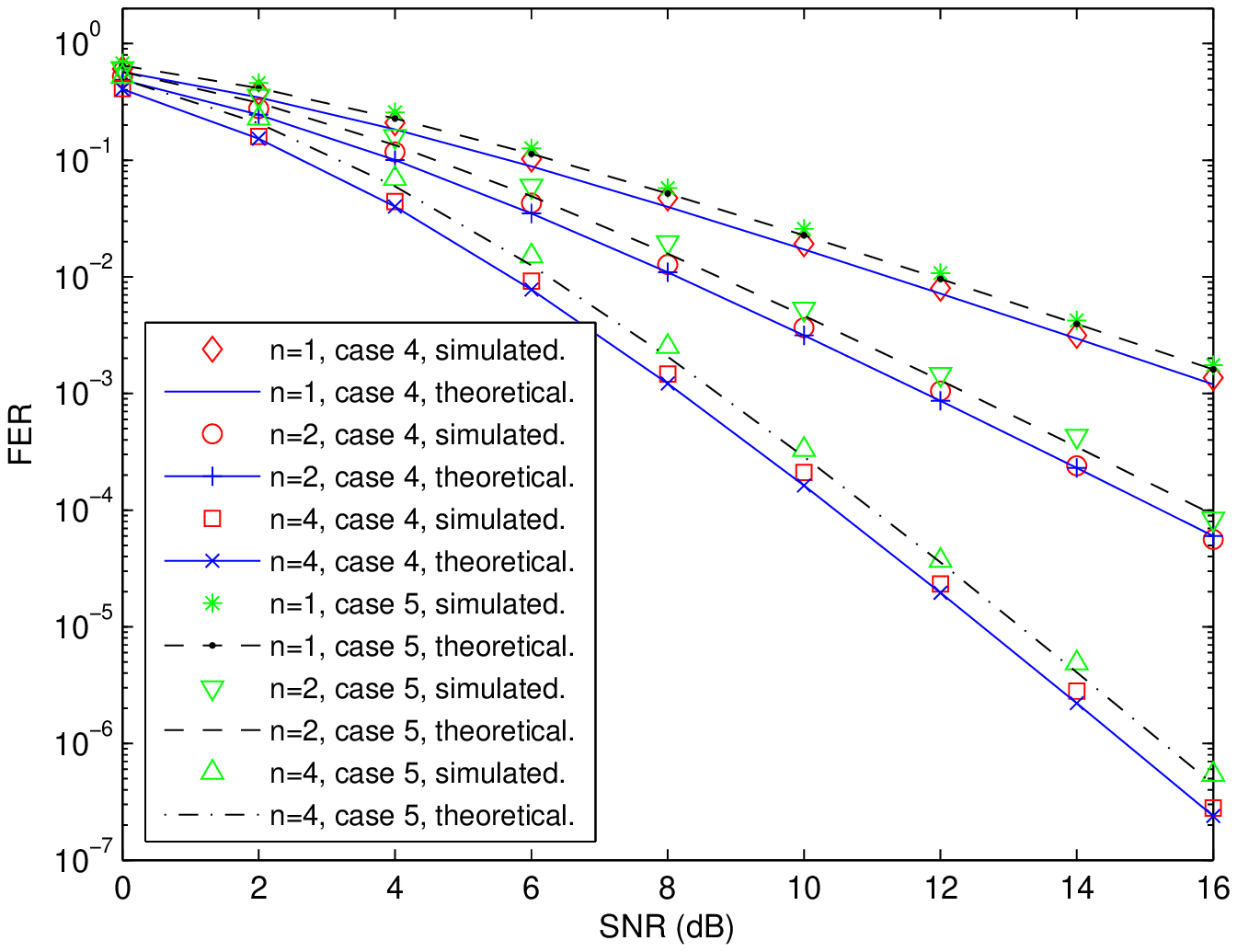}
\caption{Theoretical and simulated FER of the HRS scheme for case 4 and 5: $\Omega_0=\Omega_{1i}=\Omega_{2i}=1$, coded.} \label{fig.case.4.5}
\end{figure}
%

\begin{figure}[!t]
\centering
\includegraphics[width=0.8\textwidth]{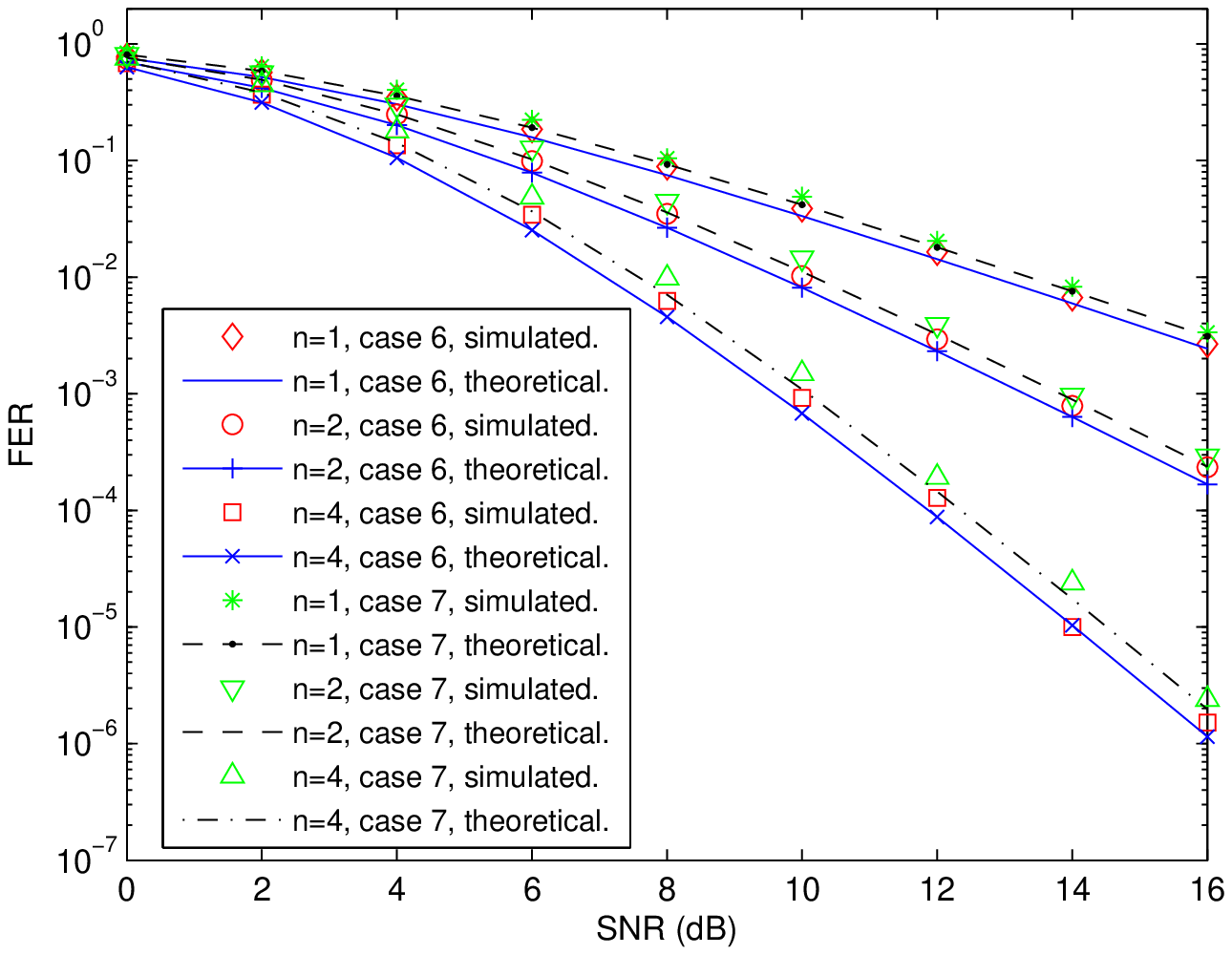}
\caption{Theoretical and simulated FER of the HRS scheme for case 6 and 7: $\Omega_0=\Omega_{1i}=\Omega_{2i}=1$, coded.} \label{fig.case.6.7}
\end{figure}



%


\begin{figure}[!t]
\centering
\includegraphics[width=0.8\textwidth]{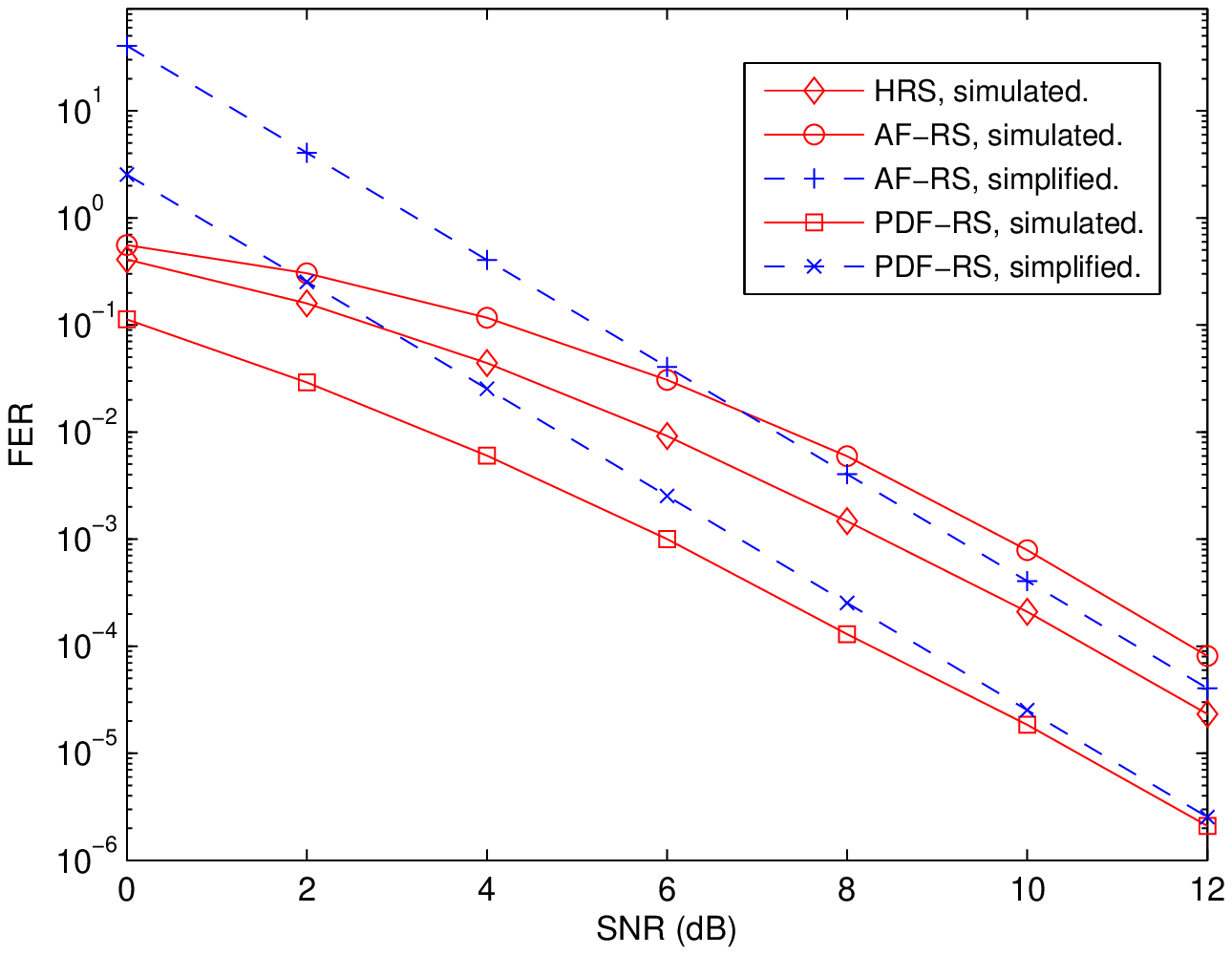}
\caption{Average FER of the HRS, AF-RS and PDF-RS schemes for case 4 with $n=4$: $\Omega_0=\Omega_{1i}=\Omega_{2i}=1$, coded.} \label{fig.case.4.AF.PDF.N4}
\end{figure}

In this section, we compare the proposed approximate FER expressions with the exact FER obtained by Monte-Carlo simulations. Unless specifically mentioned, the simulations are performed for a BPSK modulation and a frame size of $100$ or $200$ symbols over block Rayleigh fading channels. We consider
two basic cases: (1) the \emph{uncoded} case, where none channel code is used; and (2) the \emph{coded} case, where a systematic convolutional code with a code rate of $1/2$ and the generator matrix of $(5,7)_8$ or a code rate of $2/3$ and the generator matrix of $(23,35,0; 0,5,13)_8$ is used.

It is practically infeasible to verify the accuracy of analytical results for all the possible scenarios, as a prohibitively large number
of combinations can be generated by varying relay number $n$ and average SNR parameters $\Omega$. Therefore, the scenarios in Table \ref{tb.scenarios} are chosen such that a range of diverse cases are covered. For instance, cases 1, 2, and 3 are uncoded and case 4, 5, 6 and 7 are coded. Cases 1, 4, 5, and 6 are symmetric situations, while cases 2 and 3 are dissimilar situations. In all cases, 1, 2 and 4 relay nodes are considered. We also choose case 0 for comparing our proposed FER approximation model to the results of \cite{Chatzigeorgiou2008}. And in all cases, $\gamma_{t,d}=\gamma_{t,n+1}$. The SNR threshold values are obtained using the methods developed in Section \ref{sec.FER.approx.model}, and are shown in Table \ref{tb.snr.threshold}.


\begin{table*}[!t]
\centering
\caption{SCENARIOS} \label{tb.scenarios}
\begin{tabular}{| c | c| c | c| c| c| c|}
\hline
           & System   &Frame & uncoded/ &Code  & Number of  & \multirow{2}*{SNRs} \\
          &  Type     &Size  & coded    & Rate  &Nodes      &  \\  \hline \hline
case 0 & MIMO&100 & uncoded &-&  N$_T$=1, N=N$_R$=1,2,4 & $\Omega=1$\\ \hline
case 1 & HRS &100&  uncoded  &-& n=1, 2, 4 & $\Omega_0=\Omega_{1i}=\Omega_{2i}=1$ \\ \hline
case 2 & HRS &100& uncoded &-&  n=1, 2, 4 & $\Omega_0=1$,  $\Omega_{1i}=16$, $\Omega_{2i}=1$  \\ \hline
case 3 & HRS &100 & uncoded &-& n= 1, 2, 4 & $\Omega_0=1$,  $\Omega_{1i}=1/16$, $\Omega_{2i}=1$ \\ \hline
case 4 & HRS &100 & coded   &1/2& n= 1, 2, 4& $\Omega_0=\Omega_{1i}=\Omega_{2i}=1$ \\ \hline
case 5 & HRS &200 & coded   &1/2&  n= 1, 2, 4& $\Omega_0=\Omega_{1i}=\Omega_{2i}=1$ \\ \hline
case 6 & HRS &100 & coded   &2/3& n= 1, 2, 4& $\Omega_0=\Omega_{1i}=\Omega_{2i}=1$ \\ \hline
case 7 & HRS &200 & coded   &2/3&  n= 1, 2, 4& $\Omega_0=\Omega_{1i}=\Omega_{2i}=1$ \\ \hline
\end{tabular}
\end{table*}

\begin{table*}[!t]
\centering
\caption{SNR Threshold} \label{tb.snr.threshold}
\begin{tabular}{| c | c| c | c| c| c|}
\hline
 & System Type  & uncoded/coded & Number of Nodes & Diversity Order & SNR Threshold \\  \hline \hline
 \multirow{3}*{case 0} &  \multirow{3}*{MIMO} &  \multirow{3}*{uncoded }& N$_T$=1, N=N$_R$=1 & 1 & 5.10 dB\\ \cline{4-6}
 & & &N$_T$=1, N=N$_R$=2 & 2 & 5.36 dB \\ \cline{4-6}
& & & N$_T$=1, N=N$_R$=4 & 4 & 5.89 dB \\ \hline
\multirow{3}*{case 1,2,3} &  \multirow{3}*{HRS} &  \multirow{3}*{uncoded }& n=1 & 2 & 5.36 dB\\ \cline{4-6}
 & & &n=2 & 3 & 5.62 dB \\ \cline{4-6}
& & & n=4 & 5 & 6.16 dB \\ \hline
\multirow{3}*{case 4} &  \multirow{3}*{HRS} &  \multirow{3}*{coded  }& n=1 & 2 & -1.48 dB\\ \cline{4-6}
 & & &n=2 & 3 & -1.26 dB \\ \cline{4-6}
& & & n=4 & 5 & -0.81 dB \\ \hline
\multirow{3}*{case 5} &  \multirow{3}*{HRS} &  \multirow{3}*{coded  }& n=1 & 2 & -0.85 dB\\ \cline{4-6}
 & & &n=2 & 3 & -0.68 dB \\ \cline{4-6}
& & & n=4 & 5 & -0.33 dB \\ \hline
\multirow{3}*{case 6} &  \multirow{3}*{HRS} &  \multirow{3}*{coded  }& n=1 & 2 & 0.06 dB\\ \cline{4-6}
 & & &n=2 & 3 & 0.20 dB \\ \cline{4-6}
& & & n=4 & 5 & 0.47 dB \\ \hline
\multirow{3}*{case 7} &  \multirow{3}*{HRS} &  \multirow{3}*{coded  }& n=1 & 2 & 0.60 dB\\ \cline{4-6}
 & & &n=2 & 3 & 0.70 dB \\ \cline{4-6}
& & & n=4 & 5 & 0.90 dB \\ \hline
\end{tabular}
\end{table*}

Fig. \ref{fig.case.0} shows the analytical FER curves by using the proposed model and the model of  \cite{Chatzigeorgiou2008} for case 0. For case 0, the SNR threshold was found to be $4.6$ dB based on Eq. (\ref{eq.SNR.threshod.old}) (the model of \cite{Chatzigeorgiou2008}), and was found to be $5.10$ dB, $5.36$ dB and $5.89$ dB for $N=1, 2, 4$, respectively,   by our proposed model based on Eq. (\ref{eq:P_f_G_uncode}).  As shown in the figure, our proposed model converge with the simulated FER quickly as the SNR increases while the relative error using  \cite{Chatzigeorgiou2008} cannot be ignored even at high SNR.   The FER results based on  the model in \cite{Chatzigeorgiou2008} become less accurate when the diversity order increaeses, but our proposed model is still accurate. While not shown here, similar trends can be observed for other scenarios.

Fig. \ref{fig.case.1}  shows the proposed approximate FER results and simulated results of the HRS scheme for case 1. From the figure we can see that the analytical theoretical curves obtained by Eq.~(\ref{eq:FER_HRS}) converge well with the simulated results, and the simplified results computed by Eq.~(\ref{eq:FER_HRS.simplify}) converge to the simulated results at high SNRs.
Figs \ref{fig.case.4.5} and \ref{fig.case.6.7} show the different encoders and other block lengths for coded scheme and confirm the good performance of the proposed model for different parameters.
Case 2 and case 3 are also verified through simulation but omitted here for brevity.

Fig. \ref{fig.case.4.AF.PDF.N4} shows our proposed simplified FER of the AF-RS and PDF-RS for case 4 with 4 relay nodes. From the figures we
can see that  the simplified results converge to the simulated results at high SNRs for both AF-RS and PDF-RS. And the HRS scheme has considerable  performance gain over AF-RS.

\section{Conclusions} \label{sec.conclusion}

In this paper, we have analyzed the average FER of the HRS scheme in general cooperative wireless networks with and without
applying channel coding at transmitting nodes.   We proposed an improved SNR threshold-based FER approximation model. We then apply this model to HRS system and derived the analytical approximate average FER expression and the simplified asymptotic FER expression at high SNRs for the HRS scheme. Simulation results match well with the theoretical analysis, which validates our derived FER expressions.

%
\appendices
\section{The PDF and CDF of $\gamma_{HRS}$: The Proof of Eq. (\ref{eq:CDF_SNR_Z})} \label{Proof:CDF_SNR_Z}
According to Eq. (\ref{eq:gamma_vz}), $\gamma_{HRS} = \gamma_{0} + \gamma_m$. In the following,  we first derive the PDF of
$\gamma_0$ and $\gamma_m$  and then get the PDF of $\gamma_{HRS}
$ as the convolution of the PDF of $\gamma_0$ and $\gamma_m$.

As $\gamma_0$, $\gamma_{1i}$ and $\gamma_{2i}$ are exponentially
distributed, their PDF and CDF are given as
\begin{equation*}
f_{\gamma_k}(\gamma) = \lambda_{k} e^{- \lambda_{k} \gamma }, k \in
\{0,1i,2i\},
\end{equation*}
and
\begin{equation*}
F_{\gamma_k}(\gamma) = 1 - e^{- \lambda_{k} \gamma }, k \in
\{0,1i,2i\},
\end{equation*}
respectively.

According to Eq. (\ref{eq:SNR_i_cases}) and Eq. (\ref{eq:Selection}), $\gamma_m = \max_{1 \leq i \leq n} \{ \gamma_i \}$ and
\begin{equation*}
\gamma_i = \begin{cases}
\frac{\gamma_{1i}\gamma_{2i}}{\gamma_{1i}+\gamma_{2i}+1}, & \text{if $z_i=1$, }\\
\gamma_{2i}, & \text{if $z_i=0$. }
\end{cases}
\end{equation*}
To get the distribution of $\gamma_m$, we first derive the distribution of $\gamma_i$.

If $z_i=0$, $\gamma_i=\gamma_{2i}$. Then, the conditional CDF of
$\gamma_i$ given $z_i=0$ is
\begin{equation} \label{eq:CDF_gamma_i_zi0}
F_{\gamma_i}(\gamma|z_i=0) = Pr(\gamma_{2i}<\gamma) =
1-e^{-\lambda_{2i} \gamma}.
\end{equation}

If $z_i=1$, using the approximation $\frac{x y}{x+y+1} \approx min\{x,y\}$ \cite{Anghel2004}, then the conditional CDF of $\gamma_{i}$ given $z_i=1$ is \cite{David2003}.
\begin{equation} \begin{split} \label{eq.CDF.gamma_i_z1}
F_{\gamma_i|_{z_i=1}}(\gamma)
&= Pr(min\{\gamma_{1i}|_{z_i=1},\gamma_{2i}\} < \gamma) \\
&=1-(1-F_{\gamma_{1i}|_{z_i=1}}(\gamma))(1-F_{\gamma_{2i}}(\gamma)).
\end{split} \end{equation}

The conditional PDF and CDF of $\gamma_{1i}|_{z_i=1}$ are
\begin{equation*} \begin{split}
f_{\gamma_{1i}|_{z_i=1}}(\gamma) &=
\frac{f_{\gamma_{1i}}(\gamma) Pr(\gamma_{1i}=\gamma, z_i=1)}{Pr(z_i=1)}\\
&= \frac{\lambda_{1i} e^{-\lambda_{1i} \gamma} P_{f,\gamma_{1i}}^G(\gamma)}{P_{f,\gamma_{1i}}}\\
&\approx
\begin{cases}
\frac{\lambda_{1i} e^{-\lambda_{1i} \gamma} }
{1-e^{-\lambda_{1i} \gamma_{t,1}}}, & \text{if } \gamma < \gamma_{t,1}, \\
0,& \text{if } \gamma \geq \gamma_{t,1},
\end{cases}
\end{split} \end{equation*}
\begin{equation}\label{eq.CDF.gamma_1i_z1}
F_{\gamma_{1i}|_{z_i=1}}(\gamma) =
\begin{cases}
\frac{ 1-e^{-\lambda_{1i} \gamma}}
{1-e^{-\lambda_{1i} \gamma_{t,1}}}, & \text{if } \gamma < \gamma_{t,1}, \\
1,& \text{if } \gamma \geq \gamma_{t,1},
\end{cases}
\end{equation}
where $P_{f,\gamma_{1i}}^G(\gamma)$ is the FER of $\gamma_{1i}$ at AWGN channel and $P_{f,\gamma_{1i}}^G(\gamma) \approx 1$ if $\gamma < \gamma_{t,1}$ and $P_{f,\gamma_{1i}}^G(\gamma) \approx 0$ if $\gamma \geq \gamma_{t,1}$.

Combining Eq. (\ref{eq.CDF.gamma_i_z1}) and Eq. (\ref{eq.CDF.gamma_1i_z1}), the corresponding conditional CDF of $\gamma_i$ given $z_i=1$ can be written
as
\begin{equation} \begin{split} \label{eq:CDF_gamma_i_zi1}
F_{\gamma_i|z_i=1}(\gamma)
& \approx
\begin{cases}
1-\frac{ e^{-\lambda_{1i} \gamma}-e^{-\lambda_{1i} \gamma_{t,1}}}
{1-e^{-\lambda_{1i} \gamma_{t,1}}} e^{-\lambda_{2i} \gamma}, & \text{if } \gamma < \gamma_{t,1}, \\
1,& \text{if } \gamma \geq \gamma_{t,1},
\end{cases} \\
\end{split} \end{equation}
where at the second step, we extend the range of $\gamma$ from $\gamma_{t,1}$ to $\gamma_{t,d}$ using an exponential function to approximate $F_{\gamma_i|z_i=1}(\gamma)$. The simulation results in Section \ref{sec.simulation} indicate that this approximation is accurate.


Combining Eq. (\ref{eq:CDF_gamma_i_zi0}) and Eq.
(\ref{eq:CDF_gamma_i_zi1}) and noting  that $Pr(z_i=1)=1-e^{-\lambda_{1i}\gamma_{t,1}}$ and $Pr(z_i=0)=e^{-\lambda_{1i}\gamma_{t,1}}$, the CDF of $\gamma_i$ can be rewritten
as
\begin{equation} \begin{split} \label{eq:CDF_gamma_i}
F_{\gamma_i}(\gamma)
&= Pr(z_i=1)F_{\gamma_i}(\gamma|z_i=1) + Pr(z_i=0)F_{\gamma_i}(\gamma|z_i=0)\\
&\approx
\begin{cases}
1-e^{-(\lambda_{1i} + \lambda_{2i}) \gamma}, & \text{if } \gamma < \gamma_{t,1}, \\
1-e^{-(\lambda_{1i}\gamma_{t,1} + \lambda_{2i}\gamma) },& \text{if } \gamma \geq \gamma_{t,1},
\end{cases}\\
&\approx
\begin{cases}
1-e^{-(\frac{\gamma_{t,1}}{\gamma_{t,d}}\lambda_{1i} + \lambda_{2i}) \gamma}, & \text{if } \gamma < \gamma_{t,d}, \\
1-e^{-(\lambda_{1i}\gamma_{t,1} + \lambda_{2i}\gamma) },& \text{if } \gamma \geq \gamma_{t,d},
\end{cases}
\end{split} \end{equation}
Noting that $\gamma_{t,1}\lessapprox \gamma_{t,d}$, we can use a exponential function to approximate the CDF of $\gamma_i$ when $\gamma < \gamma_{t,d}$. In order to keep the value $F_{\gamma_i}(\gamma_{t,d})$ unchanged, we have  the last step.
Then the CDF of $\gamma_m$ (when $\gamma < \gamma_{t,d}$) is given by \cite{David2003}
\begin{equation} \label{eq:CDF_SNR_Max_1}
\begin{split}
F_{\gamma_m}(\gamma) &= Pr(\max_{1\leq i \leq n}\{\gamma_i\} <
\gamma)\\
&\approx \prod_{i=1}^n \left(1 -e ^{-(\frac{\gamma_{t,1}}{\gamma_{t,d}}\lambda_{1i} + \lambda_{2i})
\gamma}\right)\\
&= {\sum_{\mathbf{b} \in \mathcal{Z}_n} \mathcal{C}_1e ^{-
\mathcal{C}_2 \gamma} },
\end{split}
\end{equation}
with the help of that
\begin{equation} \label{eq.binomial}
\prod_{i=1}^n (1 + x_i) =\prod_{i=1}^n {x_i}^{0} +
x_1^1\prod_{i=2}^n {x_i}^{0} + \cdots + \prod_{i=1}^n {x_i}^{1} =
\sum_{\mathbf{b} \in \mathcal{Z}_n} \prod_{i=1}^n {x_i}^{b_i},
\end{equation}
where $\mathbf{b}=\{b_i, i=1,\cdots,n\} \in \mathcal{Z}_n, b_i \in \{0,1\}$, $\mathcal{C}_1 = (-1)^{\sum_{i=1}^n
b_i} $ and $
\mathcal{C}_2 = \sum_{i=1}^n b_i (\frac{\gamma_{t,1}}{\gamma_{t,d}}\lambda_{1i} + \lambda_{2i}) $ .

The PDF of $\gamma_m$ is
given by
\begin{equation*}
f_{\gamma_m}(\gamma) \approx -{\sum_{\mathbf{b} \in \mathcal{Z}_n, \mathbf{b} \neq 0}
\mathcal{C}_1 \mathcal{C}_2 e ^{- \mathcal{C}_2 \gamma} }.
\end{equation*}
As $\gamma_{HRS}=\gamma_0+\gamma_m$, the PDF of $\gamma_{HRS}$ when
$\gamma \leq \gamma_{t,d}$ can be expressed as
\begin{equation} \begin{split} \label{eq:PDF_gamma_z}
f_{\gamma_{HRS}}(\gamma) &= \int_0^{\gamma} f_{\gamma_0}(\gamma-t) f_{\gamma_m}(t) \text{d} t \\
&=- \int_0^{\gamma}  \lambda_0 e^{- \lambda_{0} (\gamma-t)}{\sum_{\mathbf{b} \in \mathcal{Z}_n} \mathcal{C}_1 \mathcal{C}_2 e ^{- \mathcal{C}_2 t} } \text{d} t \\
&\approx -  \lambda_0 e^{- \lambda_{0} \gamma} {\sum_{\mathbf{b} \in \mathcal{Z}_n, \mathbf{b}\neq0} \mathcal{C}_1 \mathcal{C}_2 \int_0^{\gamma} e ^{- (\mathcal{C}_2 - \lambda_0)t} } \text{d} t \\
&= -  \lambda_0 e^{- \lambda_{0} \gamma}\sum_{\mathbf{b} \in \mathcal{Z}_n, \mathbf{b}\neq0} \mathcal{C}_1
\mathcal{C}_2
\frac{1-e ^{- (\mathcal{C}_2 - \lambda_0)\gamma} }{\mathcal{C}_2 - \lambda_0} \\
&= \sum_{\mathbf{b} \in \mathcal{Z}_n, \mathbf{b}\neq0}  \mathcal{C}_1
\lambda_0 \mathcal{C}_2
\frac{e^{- \lambda_{0} \gamma}-e ^{- \mathcal{C}_2\gamma} }{
\lambda_0 -\mathcal{C}_2}.
\end{split} \end{equation}
Finally, the CDF of $\gamma_{HRS}$ is given by
\begin{equation} \begin{split} \label{eq:CDF_gamma_z}
F_{\gamma_{HRS}}(\gamma_{t,d}) &= \int_0^{\gamma_{t,d}} f_{\gamma_{HRS}}(t)\text{d}t \\
&\approx \int_0^{\gamma_{t,d}} \sum_{\mathbf{b} \in \mathcal{Z}_n, \mathbf{b}\neq0}
\mathcal{C}_1 \lambda_0 \mathcal{C}_2
\frac{e^{- \lambda_{0} t}-e ^{- \mathcal{C}_2t} }{ \lambda_0-\mathcal{C}_2} \text{d}t \\
&= \sum_{\mathbf{b} \in \mathcal{Z}_n, \mathbf{b}\neq0}
\mathcal{C}_1
\frac{\mathcal{C}_2(1-e^{- \lambda_{0} \gamma_{t,d}})-\lambda_0(1-e ^{-
\mathcal{C}_2 \gamma_{t,d}}) }{\lambda_0-\mathcal{C}_2}.
\end{split} \end{equation}
%
\section{Simplified CDF of $\gamma_{HRS}$ at high SNR:  The Proof of Eq. (\ref{eq:CDF_SNR_Z_approx})} \label{Proof:CDF_SNR_Z_approx}
 Using Eq. (\ref{eq:PDF_gamma_z}) and the
approximation that $e^{-x} \approx 1- x$, the PDF of $\gamma_{HRS}$
at high SNR can be approximated as
\begin{equation} \begin{split} \label{eq:PDF_gamma_z_a0}
f_{\gamma_{HRS}}(\gamma) &\approx \sum_{\mathbf{b} \in
\mathcal{Z}_n} \mathcal{C}_1 \lambda_0 \mathcal{C}_2
\frac{(1- \lambda_{0} \gamma)-(1- \mathcal{C}_2\gamma) }{ \lambda_0 -\mathcal{C}_2}\\
&=- \lambda_0  \sum_{\mathbf{b} \in \mathcal{Z}_n} {
\mathcal{C}_1 \mathcal{C}_2 \gamma}.
\end{split}
\end{equation}
Applying the approximation that $e^{-x} \approx 1- x$ to Eq.
(\ref{eq:CDF_gamma_i}) and Eq. (\ref{eq:CDF_SNR_Max_1}), we can get
\begin{equation} \label{eq:CDF_gamma_m_1}
F_{\gamma_m}(\gamma) \approx  \gamma^n \prod_{i=1}^n (\frac{\gamma_{t,1}}{\gamma_{t,d}}\lambda_{1i} + \lambda_{2i}),
\end{equation}
and
\begin{equation} \label{eq:CDF_gamma_m_2}
F_{\gamma_m}(\gamma) \approx \sum_{\mathbf{b} \in \mathcal{Z}_n}
\mathcal{C}_1 (1-\mathcal{C}_2\gamma) = \sum_{\mathbf{b} \in
\mathcal{Z}_n} \mathcal{C}_1 -\sum_{\mathbf{b} \in \mathcal{Z}_n}
\mathcal{C}_1 \mathcal{C}_2 \gamma.
\end{equation}

As $F_{\gamma_m}$ is CDF, $F_{\gamma_m}(0) = 0$, we can get
\begin{equation} \label{eq:CDF_gamma_m_0}
F_{\gamma_m}(0) = \sum_{\mathbf{b} \in \mathcal{Z}_n} \mathcal{C}_1
= 0.
\end{equation}
Combining Eq. ({\ref{eq:CDF_gamma_m_1}}), Eq.
(\ref{eq:CDF_gamma_m_2}), and Eq. (\ref{eq:CDF_gamma_m_0}), we can
get
\begin{equation} \label{eq:CDF_gamma_m_3}
-\sum_{\mathbf{b} \in \mathcal{Z}_n} \mathcal{C}_1 \mathcal{C}_2
\gamma \approx  \gamma^n \prod_{i=1}^n (\frac{\gamma_{t,1}}{\gamma_{t,d}}\lambda_{1i} + \lambda_{2i}).
\end{equation}

Substituting Eq. (\ref{eq:CDF_gamma_m_3}) into Eq.
(\ref{eq:PDF_gamma_z_a0}), we can get the approximated PDF of
$\gamma_{HRS}$ as
\begin{equation*}
f_{\gamma_{HRS}}(\gamma) \approx \lambda_0 \gamma^n \prod_{i=1}^n (\frac{\gamma_{t,1}}{\gamma_{t,d}}\lambda_{1i} + \lambda_{2i}),
\end{equation*}

And the CDF of $\gamma_{HRS}$ can be expressed
\begin{equation*} \begin{split} \label{eq:CDF_gamma_z_a}
F_{\gamma_{HRS}}(\gamma_{t,d})
= \int_0^{\gamma_{t,d}} f_{\gamma_{HRS}}(t)  \text{d} t
\approx \frac{\lambda_0 \gamma_{t,d}^{n+1}}{n+1} \prod_{i=1}^n
(\frac{\gamma_{t,1}}{\gamma_{t,d}}\lambda_{1i} + \lambda_{2i}).
\end{split}\end{equation*}

\bibliographystyle{ieeetran}
\bibliography{IEEEabrv,HRS}

\end{document}